\documentclass[aps,nofootinbib,twocolumn]{revtex4}%
\usepackage{amsfonts}
\usepackage{amsmath}
\usepackage{amssymb}
\usepackage{color}
\usepackage{graphicx}%
\providecommand{\U}[1]{\protect\rule{.1in}{.1in}}

\begin{document}
\preprint{BL10952}
\title[ ]{Many-body large polaron optical conductivity in SrTi$_{1-x}$Nb$_{x}$O$_{3}$}
\author{J.~T.~Devreese$^{1,\ast}$}
\author{S.~N.~Klimin$^{1,\ast\ast}$}
\author{J.~L.~M.~van~Mechelen$^{2}$}
\author{D.~van~der~Marel$^{2}$}
\affiliation{$^{1}$Theorie van Kwantumsystemen en Complexe Systemen, Universiteit
Antwerpen, Groenenborgerlaan 171, B-2020 Antwerpen, Belgium}
\affiliation{$^{2}$D\'{e}partement de Physique de la Mati\`{e}re Condens\'{e}e,
Universit\'{e} de Gen\`{e}ve, Gen\`{e}ve, Switzerland}

\begin{abstract}
Recent experimental data on the optical conductivity of niobium doped
SrTiO$_{3}$ are interpreted in terms of a gas of large polarons with effective
coupling constant $\alpha_{eff}\approx2$. The {theoretical approach takes into
account} many-body effects, the electron-phonon interaction with multiple
LO-phonon branches, and the degeneracy and the anisotropy of the {Ti t$_{2g}$}
conduction band. {Based on the Fr\"{o}hlich interaction, the many-body
large-polaron theory} provides an interpretation for the essential
characteristics, except -- interestingly -- for the unexpectedly large
intensity of a peak at $\sim130$ meV, of the observed optical conductivity
spectra of SrTi$_{1-x}$Nb$_{x}$O$_{3}$ {\textit{without}} any {adjustment} of
material parameters.

\end{abstract}
\volumeyear{ }
\volumenumber{ }
\issuenumber{ }
\eid{ }
\date{\today}
\startpage{1}
\endpage{ }
\maketitle

\section{Introduction \label{sec:intro}}

The infrared optical absorption of perovskite-type materials, in particular,
{of} copper oxide based high-$T_{c}$ superconductors {and of the manganites}
has been the subject of intensive investigations
\cite{Lupi1999,calva0,falck1,calva1b,QQ4,Crawford90,zhang,Homes1997,Ronnow,Hartinger}%
. {Insulating SrTiO$_{3}$ has a perovskite structure and manifests a
metal--insulator transition at room temperature around a doping of 0.002\% La
or Nb per unit cell \cite{Calvani1993}. At low doping concentrations, between
0.003\% and 3\%, strontium titanate reveals a} superconducting phase
transition \cite{Schooley1964} below 0.7 K. {Various optical experiments
\cite{Gervais93,Calvani1993,Eagles96,Ang2000,JPCM2006,VDM-PRL2008} show a
mid-infrared band in the normal state optical conductivity of doped
SrTiO$_{3}$ which is often explained by polaronic behavior.}
In the recently observed optical conductivity spectra of
Ref.~\cite{VDM-PRL2008}, shown in Fig.~1, there is a broad mid-infrared
optical conductivity band starting at a photon energy of $\hbar\Omega\sim100$
meV, which is within the range of the LO-phonon energies of SrTi$_{1-x}%
$Nb$_{x}$O$_{3}$. The peaks/shoulders of the experimental optical conductivity
band at $\hbar\Omega\sim200$ to 400 meV resemble the peaks provided by the
mixed plasmon-phonon excitations as described in Ref.~\cite{TD2001}. Based on
the experimental data, the authors deduce a coupling constant $3<\alpha<4$ and
conclude the mid-infrared peaks to originate from large polaron formation. The
high and narrow peaks positioned at the lower frequencies with respect to the
mid-infrared band are attributed in Ref.~\cite{VDM-PRL2008} to the optical
absorption of the TO-phonons.

\begin{figure}
\hspace{-1.7mm}\includegraphics[width=7cm]{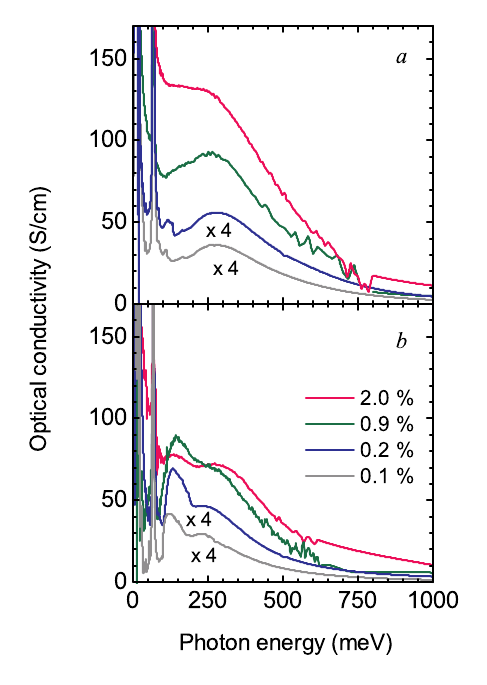}
\caption{ Optical conductivity of SrTi$_{1-x}$Nb$_{x}$O$_{3}$ for
0.1\% (grey curves), 0.2\% (blue curves), 0.9\% ({green} curves) and 2\% (pink
curves) at 300 K (panel {a}) and 7 K (panel {b}). For clarity, the
mid-infrared conductivities of $x=0.1\%$ and 0.2\% are magnified by {a} factor
4. (From Ref.~\cite{VDM-PRL2008}.)}
\label{fig_swtransfer}
\end{figure}

There are different types of polaron states in solids. In the effective mass
approximation for the electron placed in a continuum polarizable medium, a
so-called large or continuum polaron can exist. Large polaron wave functions
and the corresponding lattice distortions spread over many lattice sites. Due
to the finite phonon frequencies the ion polarizations can follow the polaron
motion if the motion is sufficiently slow. Hence, large polarons with a low
kinetic energy propagate through the lattice as free electrons but with an
enhanced effective mass. When the polaron binding energy is larger than the
half bandwidth of the electron band, all states in the Bloch bands are
`dressed' by phonons. In this strong-coupling regime, the finite electron
bandwidth becomes important, so the continuum approximation cannot be applied.
In this case the carriers are described as \textquotedblleft
small\textquotedblright\ or discrete (lattice) polarons that can hop between
different states localized at lattice sites. A key distinction between large
and small polarons is then the radius of the polaron state. For large
polarons, that radius substantially exceeds the lattice constant, while for
small polarons it is comparable to the lattice constant. A review of the
properties of large and small polarons can be found, e.~g., in Refs.
\cite{Review2009,Book}. In the theory of \textquotedblleft
mixed\textquotedblright\ polarons \cite{E1,E2,E3,Eagles85} the states of the
electron-phonon system are composed of a mixture of large and small polaron states.

Polaron states are formed due to the electron-phonon interaction, which is
different in the cases of large and small polarons. For a large polaron, the
electron-phonon interaction is provided by a macroscopic (continuum)
polarization of the lattice. This interaction is characterized by the coupling
constant $\alpha$ introduced by Fr\"{o}hlich \cite{Fr1954},%
\begin{equation}
\alpha=\frac{1}{2}\left(  \frac{1}{\varepsilon_{\infty}}-\frac{1}%
{\varepsilon_{0}}\right)  \frac{e^{2}}{\hbar\omega_{L}}\left(  \frac
{2m_{b}\omega_{L}}{\hbar}\right)  ^{1/2}, \label{alpha}%
\end{equation}
where $\varepsilon_{\infty}$ and $\varepsilon_{0}$ are, respectively, the
high-frequency and the static dielectric constants, $e$ is the electron
charge, $\omega_{L}$ is the longitudinal optical (LO) phonon frequency in the
Brillouin zone centre, and $m_{b}$ is the band electron (or hole) mass. The
large-polaron coupling constant is thus expressed through macroscopic
observable parameters of a polarizable medium. On the contrary, the
electron-phonon interaction for a small polaron is described through
microscopic parameters.

The nature of the polaron states in SrTi$_{1-x}$Nb$_{x}$O$_{3}$ is not yet
clear. Previous optical measurements on strontium titanate were interpreted in
terms of small polarons \cite{Reik67,Eagles85}. However, that assumption
contradicts the interpretation of transport measurements \cite{Frederikse},
which rather support the large-polaron picture. Also the heat capacity
measurements \cite{Ambler66}, provide effective masses similar to those of
large polarons. In Ref.~\cite{Eagles96}, the experimental results of Ref.
\cite{Gervais93} on the temperature-dependent plasma frequencies in
SrTi$_{1-x}$Nb$_{x}$O$_{3}$ were interpreted within the theory of mixed
polarons \cite{E1,E2,E3,Eagles85}. Thermoelectric power measurements
\cite{Frederikse} have shown that the density-of-states masses increase with
increasing temperature, which can be explained by a theory of mixed polarons
\cite{E1}. It has been supposed \cite{JPCM2006} that the polaron optical
conductivity in SrTi$_{1-x}$Nb$_{x}$O$_{3}$ is probably provided by mixed
polarons. A possible coexistence of large and small mass polarons has been
suggested in Ref. \cite{Iadonisi1998}. In Ref. \cite{Eagles95}, coexistence of
small and large polarons in the same solid is invoked to interpret
experimental data on the optical absorption in oxides.

The key question is to determine the type of polarons that provide the
mechanism of the polaron optical conductivity in SrTi$_{1-x}$Nb$_{x}$O$_{3}$.
The optical response of large polarons in various approximations was studied,
e.~g., in Refs. \cite{GLF,DHL1971,KED1969,DSG}. The same problem for the small
polaron was investigated in \cite{Reik67,Emin1993}. In the large-polaron
theory, the optical absorption is provided by transitions (with $0,1,\ldots$
phonon emission) between different continuum electron states. In the
small-polaron theory, the optical absorption occurs when the self-trapped
carrier is induced to transfer from its localized state to a localized state
at an adjacent site, with emission of phonons. Because of the different
physical mechanisms involved, the optical conductivity spectra of large and
small polarons are different from each other. In the large-polaron theory the
polaron optical conductivity behaves at high frequencies $\Omega$ as a power
function $\left(  \propto\Omega^{-5/2}\right)  $. In the small-polaron theory,
the polaron optical conductivity at high frequencies decreases much faster
than for large polarons: as a Gaussian exponent. Therefore the analysis of
optical measurements can shed some light on the aforesaid question on the type
of polarons responsible for the optical conductivity in SrTi$_{1-x}$Nb$_{x}%
$O$_{3}$.

The polaron optical conductivity band of SrTi$_{1-x}$Nb$_{x}$O$_{3}$ occupies
the mid-infrared range of the photon energies $\hbar\Omega\lesssim1$ eV, and
the threshold for interband electron-hole transitions lies at the band gap
energy, which is around 3.3 eV in SrTi$_{1-x}$Nb$_{x}$O$_{3}$
\cite{VDM-PRL2008}. Therefore interband transitions do not interfere with the
polaron optical conductivity. Other mechanisms of electron intraband
scattering (for example, electron-phonon interaction with acoustic phonons
and/or electron or hole transitions from impurity centers) may be manifested
together with the polaron mechanism in the energy range $\hbar\Omega\lesssim1$
eV. The treatment of those mechanisms is, however, beyond the scope of the
present investigation.

We can make some preliminary suggestions concerning the dominating mechanism
of the mid-infrared optical conductivity in the Nb doped strontium titanate.
The low-frequency edge of the mid-infrared band in SrTi$_{1-x}$Nb$_{x}$O$_{3}$
at a low temperature ($T=7$ K) lies in the range of the LO-phonon energies
obtained in \cite{Gervais93}. The maximum of the mid-infrared band lies
relatively close to this low-frequency edge (the difference in frequency
between the low-frequency edge and the maximum of the mid-infrared band is
comparable to the LO-phonon frequencies in SrTi$_{1-x}$Nb$_{x}$O$_{3}$). This
behavior is characteristic of large-polaron optical conductivity rather than
of small-polaron optical conductivity. Indeed, the maximum of the small
polaron optical conductivity band is expected to be shifted to considerably
higher frequencies with respect to the low-frequency edge of the polaron
optical conductivity band (see, e.g.,~Ref. \cite{Emin1993}). Also, at
sufficiently high frequencies, the experimental mid-infrared band from Ref.
\cite{VDM-PRL2008} decreases with increasing $\Omega$ rather slowly, which is
characteristic for large-polaron optical conductivity rather than for
small-polaron optical conductivity. We therefore can suggest that the
large-polaron picture is the most appropriate for the interpretation of the
mid-infrared band of SrTi$_{1-x}$Nb$_{x}$O$_{3}$ observed in Ref.
\cite{VDM-PRL2008}.

In order to interpret the mid-infrared band of the experimental optical
conductivity spectra of SrTi$_{1-x}$Nb$_{x}$O$_{3}$ \cite{VDM-PRL2008} in
terms of polarons, we calculate the large-polaron optical conductivity spectra
for SrTi$_{1-x}$Nb$_{x}$O$_{3}$ using the model for the optical conductivity
of a large-polaron gas developed in Ref. \cite{TD2001}, adapted to take into
account multiple LO-phonon branches \cite{draft}. The degeneracy and the
anisotropy of the conduction band in SrTi$_{1-x}$Nb$_{x}$O$_{3}$ are taken
into account.

The paper is organized as follows. In Sec. \ref{sec:theory} we describe the
theoretical formalism for the polaron optical conductivity. In Sec.
\ref{sec:results} the calculated polaron optical conductivity spectra are
discussed and compared with the experiment \cite{VDM-PRL2008}. Sec.
\ref{sec:conclusions} contains conclusions.

\section{Optical conductivity of a gas of large polarons \label{sec:theory}}

The optical absorption spectra of SrTi$_{1-x}$Nb$_{x}$O$_{3}$ are sensitive to
the doping level \cite{VDM-PRL2008}. Therefore a many-polaron description is
in order. In our context, \textquotedblleft many-polaron
description\textquotedblright\ means an account of many-electron effects on
the optical conductivity of a polaron gas. These effects include the influence
of the electron-electron Coulomb interaction (which leads to screening
effects) and of the Fermi statistics of the polaron gas on the optical
conductivity spectra. In the low-density limit, those many-body effects are
not important, and the optical conductivity of a polaron gas is well described
by the optical conductivity of a single polaron multiplied by the electron
density. The scope of the present study embraces a wide range of electron
densities for which the single-polaron approach is, in general, insufficient.
As shown below, even at the lowest electron density involved in the experiment
\cite{VDM-PRL2008}, the shape and magnitude of the optical conductivity
spectrum is strongly affected by many-body effects.

We wish to compare the experiments of Ref. \cite{VDM-PRL2008}, in particular
the observed mid-infrared band, to the theoretical optical conductivity of a
gas of large polarons. For that purpose we use the many-body large polaron
approach of Refs. \cite{TD2001,draft}, which takes into account the
electron-electron interaction and the Fermi statistics of polarons.

Refs. \cite{TD2001,draft} are limited to the study of weak-coupling polarons.
Up to $\alpha\approx3$, which includes the case of SrTi$_{1-x}$Nb$_{x}$O$_{3}%
$, the weak coupling approximation can be expected to describe the main
characteristics of the many-polaron optical response (see, e.g., Refs.
\cite{TD2001,Review2009,Book}). In Ref. \cite{draft} a generalization of Ref.
\cite{TD2001} is presented that takes into account the electron-phonon
interaction with \emph{multiple LO-phonon branches} as they exist, e. g., in
complex oxides. For a single polaron, effects related to multiple LO-phonon
branches were investigated in Ref. \cite{Ferro}. The starting point for the
treatment of a many-polaron system is the Fr\"{o}hlich Hamiltonian%
\begin{align}
H  &  =\sum_{\mathbf{k}}\sum_{\sigma=\pm1/2}\frac{\hbar^{2}k^{2}}{2m_{b}%
}c_{\mathbf{k},\sigma}^{+}c_{\mathbf{k},\sigma}+\sum_{\mathbf{q}}\sum
_{j=1}^{n}\hbar\omega_{L,j}a_{\mathbf{q},j}^{+}a_{\mathbf{q},j}+U_{e-e}%
\nonumber\\
&  +\frac{1}{\sqrt{V}}\sum_{\mathbf{q}}\sum_{j=1}^{n}\left(  V_{\mathbf{q}%
,j}a_{\mathbf{q},j}\sum_{\mathbf{k}}\sum_{\sigma=\pm1/2}c_{\mathbf{k}%
+\mathbf{q},\sigma}^{+}c_{\mathbf{k},\sigma}+\mathtt{h.c.}\right)  , \label{H}%
\end{align}
where $c_{\mathbf{k},\sigma}^{+}$ ($c_{\mathbf{k},\sigma}$) are the creation
(annihilation) operators for an electron with momentum $\mathbf{k}$ and with
the spin $z$-projection $\sigma$, $a_{\mathbf{q},j}^{+}$ ($a_{\mathbf{q},j}$)
are the creation (annihilation) operators for a phonon of the $j$-th branch
with the momentum $q$, $\omega_{L,j}$ are the LO-phonon frequencies
(approximated here as non-dispersive), and $V$ is the volume of the crystal.
The polaron interaction amplitude $V_{\mathbf{q},j}$ is \cite{Ferro}%
\begin{equation}
V_{\mathbf{q},j}=\frac{\hbar\omega_{L,j}}{q}\left(  \frac{4\pi\alpha_{j}}%
{V}\right)  ^{1/2}\left(  \frac{\hbar}{2m_{b}\omega_{L,j}}\right)  ^{1/4},
\label{V2}%
\end{equation}
where $\alpha_{j}$ is a dimensionless partial coupling constant characterizing
the interaction between an electron and the $j$-th LO-phonon branch. The
electron-electron interaction is described by the Coulomb potential energy
\begin{equation}
U_{e-e}=\frac{1}{2}\sum_{\mathbf{q}\neq0}\frac{4\pi e^{2}}{\varepsilon
_{\infty}q^{2}}\sum_{\mathbf{k},\mathbf{k}^{\prime},\sigma,\sigma^{\prime}%
}c_{\mathbf{k}+\mathbf{q},\sigma}^{+}c_{\mathbf{k}^{\prime}-\mathbf{q}%
,\sigma^{\prime}}^{+}c_{\mathbf{k}^{\prime},\sigma^{\prime}}c_{\mathbf{k}%
,\sigma}. \label{Uee}%
\end{equation}

Optical phonons in SrTiO$_{3}$ show a considerable dispersion (see, e. g.,
Ref. \cite{Choudhury} and references therein). The effect of the phonon
dispersion is a broadening of features of the polaron optical conductivity
band. The magnitude of the broadening is characterized by the dispersion
parameter $\Delta\omega$ of the optical phonons, that contribute to the
integrals over $\mathbf{q}$ entering the polaron optical conductivity. In a
polar crystal with a single LO-phonon branch, that range of convergence is
approximately $q_{0}=\left(  m_{b}\omega_{LO}/\hbar\right)  ^{1/2}$. For
SrTiO$_{3}$, taking $\omega_{LO}=\max\left\{  \omega_{L,j}\right\}  $, we
obtain $q_{0}\approx1.02\times10^{9}%
\operatorname{m}%
^{-1}$. The boundary of the Brillouin zone $\pi/a_{0}$ in SrTiO$_{3}$ (where
the lattice constant $a_{0}\approx0.3905$ nm) is at $8\times10^{9}%
\operatorname{m}%
^{-1}$. Therefore the integration domain for the relevant integrals is one
order smaller than the size of the Brillouin zone. In the region $0<q<q_{0}$,
the dispersion parameter of the LO-phonon frequencies, $\Delta\omega$, is a
few percent of $\omega_{L,j}$. Consequently, $\Delta\omega$ is very small
compared with the characteristic width of the polaron band. Therefore, in the
present treatment, we apply the approximation of non-dispersive phonons.

For a description of a polarizable medium with $n$ optical-phonon branches, we
use the model dielectric function \cite{T1972,mmc3}\textrm{ }%
\begin{equation}
\varepsilon\left(  \omega\right)  =\varepsilon_{\infty}\prod_{j=1}^{n}\left(
\frac{\omega^{2}-\omega_{L,j}^{2}}{\omega^{2}-\omega_{T,j}^{2}}\right)  ,
\label{DF}%
\end{equation}
whose zeros (poles) correspond to the LO(TO) phonon frequencies $\omega_{L,j}$
($\omega_{T,j}$). This dielectric function is the result of the
straightforward extension of the Born-Huang approach \cite{BH1954} to the case
where more than one optical-phonon branch exists in a polar crystal. The
Born-Huang approach and its extension \cite{T1972} generate expressions for
the macroscopic polarization induced by the polar vibrations, and for the
corresponding electrostatic potential. This electrostatic potential is a basis
element of the Hamiltonian of the electron-phonon interaction. In Ref.
\cite{T1972}, the Hamiltonian of the electron-phonon interaction has been
explicitly derived with the amplitudes%
\begin{equation}
V_{\mathbf{q}j}=\frac{1}{\sqrt{V}}\frac{e}{iq}\left(  \frac{4\pi\hbar}{\left.
\frac{\partial\varepsilon\left(  \omega\right)  }{\partial\omega}\right\vert
_{\omega=\omega_{L,j}}}\right)  ^{1/2}. \label{rrr3}%
\end{equation}
Using Eqs. (\ref{V2}) and (\ref{rrr3}) with the dielectric function
(\ref{DF}), we arrive at the following set of linear equations for the
coupling constants $\alpha_{j}$ ($j=1,\ldots,n$):%
\begin{equation}
\sum_{k=1}^{n}\hbar\omega_{L,k}^{3}\left(  \frac{\hbar}{2m_{b}\omega_{L,k}%
}\right)  ^{1/2}\frac{\alpha_{k}}{\omega_{L,k}^{2}-\omega_{T,j}^{2}}%
=\frac{e^{2}}{2\varepsilon_{\infty}}. \label{set0}%
\end{equation}
Knowledge of the band mass, of the electronic dielectric constant
$\varepsilon_{\infty}$ and of the LO- and TO-phonon frequencies is sufficient
to determine the coupling constants $\alpha_{j}$ taking into account mixing
between different optical-phonon branches. In the particular case of a single
LO-phonon branch, Eq. (\ref{set0}) is reduced to (\ref{alpha}).

In order to describe the optical conductivity of a polaron gas, we refer to
the work \cite{PD1983}, where the Mori-Zwanzig projection operator technique
has been used to rederive the path-integral result of Ref. \cite{FHIP} and the
impedance of Ref. \cite{DSG}. We repeat the derivations of Ref. \cite{PD1983}
with the replacement of single-electron functions by their many-electron
analogs. For example, $e^{i\mathbf{q\cdot r}}$ in the Hamiltonian of the
electron-phonon interaction is replaced by the Fourier component of the
electron density for an $N$-electron system,
\begin{equation}
\rho\left(  \mathbf{q}\right)  \equiv\sum_{s=1}^{N}e^{i\mathbf{q\cdot r}_{s}%
}=\sum_{\mathbf{k},\sigma}c_{\mathbf{k}+\mathbf{q},\sigma}^{+}c_{\mathbf{k}%
,\sigma}. \label{rhoq}%
\end{equation}
As a result, we arrive at a formula which is structurally similar to the
single-polaron optical conductivity \cite{DSG,PD1983},%
\begin{equation}
\sigma\left(  \Omega\right)  =\frac{e^{2}n_{0}}{m_{b}}\frac{i}{\Omega
-\chi\left(  \Omega\right)  /\Omega}, \label{4}%
\end{equation}
where $n_{0}=N/V$ is the carrier density, and $\chi\left(  \Omega\right)  $ is
the memory function. The same many-electron derivation as in the present work,
to the best of our knowledge, was first performed for the polaron gas in 2D in
Ref. \cite{Wu1986} in the weak electron-phonon coupling limit.

In Refs. \cite{DSG,PD1983} the single-polaron memory function was calculated
starting from the all-coupling Feynman variational principle
\cite{Feynman1955}. For a many-polaron system, an effective all-coupling
extension of that variational principle has not been worked out yet. In the
present treatment, we restrict ourselves to the weak-coupling approximation
for the electron-phonon interaction to derive the memory function. In this
approximation, the memory function $\chi\left(  \Omega\right)  $ is similar to
that of Ref. \cite{Wu1986}, with two distinctions: (1) the electron gas in the
present treatment is three-dimensional, (2) several LO phonon branches are
taken into account. The resulting form of the memory function is
\begin{align}
\chi\left(  \Omega\right)   &  =\frac{4}{3\hbar m_{b}n_{0}V}\sum
_{\mathbf{q},j}q^{2}\left\vert V_{\mathbf{q},j}\right\vert ^{2}\int
_{0}^{\infty}dt\left(  e^{i\Omega t}-1\right) \nonumber\\
&  \times\operatorname{Im}\left[  \frac{\cos\left[  \omega_{L,j}\left(
t+i\hbar\beta/2\right)  \right]  }{\sinh\left(  \beta\hbar\omega
_{L,j}/2\right)  }S\left(  \mathbf{q},t\right)  \right]  , \label{xi}%
\end{align}
where $\beta=1/\left(  k_{B}T\right)  $. The dynamical structure factor
$S\left(  \mathbf{q},t\right)  $ is proportional to the two-point correlation
function (cf. Ref. \cite{TD2001}),%
\begin{equation}
S\left(  \mathbf{q},t\right)  \equiv\frac{1}{2}\left\langle \sum_{i,j=1}%
^{N}e^{i\mathbf{q}\cdot\left[  \mathbf{r}_{j}\left(  t\right)  -\mathbf{r}%
_{k}\left(  0\right)  \right]  }\right\rangle =\frac{1}{2}\left\langle
\rho\left(  \mathbf{q,}t\right)  \rho\left(  -\mathbf{q},0\right)
\right\rangle . \label{FF}%
\end{equation}
To obtain $\chi\left(  \Omega\right)  $ to order $\alpha$ it is sufficient to
perform the averaging in the correlation function (\ref{FF}) using the
Hamiltonian (\ref{H}) without the electron-phonon interaction and keeping the
electron-electron interaction term $U_{e-e}$.

We calculate the dynamical structure factor (\ref{FF}) extending the method
\cite{TD2001} to nonzero temperatures. In Ref. \cite{TD2001}, the key
advantage of the many-polaron variational approach \cite{LDB77} is exploited:
the fact that the many-body effects are entirely contained in the dynamical
structure factor $S\left(  \mathbf{q},t\right)  $. The structure factor can be
calculated using various approximations. Terms of order of $|V_{\mathbf{q}%
,j}|^{2}$ are automatically taken into account in the memory function
(\ref{xi}). Consequently, up to order $\alpha$ for $\sigma\left(
\Omega\right)  $, it is sufficient to calculate $S\left(  \mathbf{q},t\right)
$ without the electron-phonon coupling. In Ref. \cite{TD2001}, $S\left(
\mathbf{q},t\right)  $ was calculated within two different approximations: (i)
the Hartree-Fock approximation, (ii) the random-phase approximation (RPA). As
shown in Ref. \cite{TD2001}, the RPA dynamical structure factor, contrary to
the Hartree-Fock approximation, takes into account the effects both of the
Fermi statistics and of the electron-electron interaction on the many-polaron
optical-absorption spectra.

The dynamical structure factor is expressed through the density-density
Green's functions defined as%
\begin{align}
\mathcal{G}\left(  \mathbf{q},\Omega\right)   &  \equiv-i\int_{0}^{\infty
}e^{i\Omega t}\left\langle \rho\left(  \mathbf{q,}t\right)  \rho\left(
-\mathbf{q},0\right)  \right\rangle dt,\label{GF1}\\
G^{R}\left(  \mathbf{q},\Omega\right)   &  \equiv-i\int_{0}^{\infty}e^{i\Omega
t}\left\langle \left[  \rho\left(  \mathbf{q,}t\right)  ,\rho\left(
-\mathbf{q},0\right)  \right]  \right\rangle dt. \label{GF2}%
\end{align}
In terms of $G\left(  \mathbf{q},\Omega\right)  $ and $G^{R}\left(
\mathbf{q},\Omega\right)  $, the memory function (\ref{xi}) takes the form:%
\begin{align}
\chi\left(  \Omega\right)   &  =\sum_{j}\frac{\alpha_{j}\hbar\omega_{L,j}^{2}%
}{6\pi^{2}Nm_{b}}\left(  \frac{\hbar}{2m_{b}\omega_{L,j}}\right)
^{1/2}\nonumber\\
&  \times\int d\mathbf{q}\left\{  \mathcal{G}\left(  \mathbf{q},\Omega
-\omega_{L,j}\right)  +\mathcal{G}^{\ast}\left(  \mathbf{q},-\Omega
-\omega_{L,j}\right)  -\mathcal{G}\left(  \mathbf{q},-\omega_{L,j}\right)
-\mathcal{G}^{\ast}\left(  \mathbf{q},-\omega_{L,j}\right)  \right.
\nonumber\\
&  +\frac{1}{e^{\beta\hbar\omega_{L,j}}-1}\left[  G^{R}\left(  \mathbf{q}%
,\Omega-\omega_{L,j}\right)  +\left(  G^{R}\left(  \mathbf{q},-\Omega
-\omega_{L,j}\right)  \right)  ^{\ast}\right. \nonumber\\
&  \left.  \left.  -G^{R}\left(  \mathbf{q},-\omega_{L,j}\right)  -\left(
G^{R}\left(  \mathbf{q},-\omega_{L,j}\right)  \right)  ^{\ast}\right]
\right\}  . \label{MF}%
\end{align}
Taking into account the Coulomb electron-electron interaction within RPA, the
retarded Green's function $G^{R}\left(  \mathbf{q},\Omega\right)  $ is given
by%
\begin{equation}
G^{R}\left(  \mathbf{q},\Omega\right)  =\frac{\hbar VP^{\left(  1\right)
}\left(  \mathbf{q},\Omega\right)  }{1-\frac{4\pi e^{2}}{\varepsilon_{\infty
}q^{2}}P^{\left(  1\right)  }\left(  \mathbf{q},\Omega\right)  }, \label{GH}%
\end{equation}
where $P^{\left(  1\right)  }\left(  \mathbf{q},\Omega\right)  $ is the
polarization function of the free electron gas, see, e.g., \cite{Mahan}%
\begin{equation}
P^{\left(  1\right)  }\left(  \mathbf{q},\Omega\right)  =\frac{1}{V}%
\sum_{\mathbf{k},\sigma}\frac{f_{\mathbf{k}+\mathbf{q},\sigma}-f_{\mathbf{k}%
,\sigma}}{\hbar\Omega+\frac{\hbar^{2}\left(  \mathbf{k}+\mathbf{q}\right)
^{2}}{2m_{b}}-\frac{\hbar^{2}k^{2}}{2m_{b}}+i\delta},\quad\delta\rightarrow+0
\label{P1}%
\end{equation}
with the electron average occupation numbers $f_{\mathbf{k},\sigma}$. The
function $\mathcal{G}\left(  \mathbf{q},\Omega\right)  $ is obtained from
$G^{R}\left(  \mathbf{q},\Omega\right)  $ using the exact analytical relation%
\begin{equation}
\left(  1-e^{-\beta\hbar\Omega}\right)  \operatorname{Im}\mathcal{G}\left(
\mathbf{q},\Omega\right)  =\operatorname{Im}G^{R}\left(  \mathbf{q}%
,\Omega\right)  \label{pr1}%
\end{equation}
and the Kramers-Kronig dispersion relations for $\mathcal{G}\left(
\mathbf{q},\Omega\right)  $.

The above expressions are written for an isotropic conduction band. However,
the conduction band of SrTi$_{1-x}$Nb$_{x}$O$_{3}$ is strongly anisotropic and
triply degenerate. The electrons are doped in three bands: $d_{xy}$, $d_{yz}$
and $d_{xz}$, which all have their minima at $\mathbf{k}=0$. Each of these
bands has light masses along two direction ($x$ and $y$ for $d_{xy}$, etc.)
and a heavy mass along the third direction. While each electron has a strongly
anisotropic mass, the electronic transport remains isotropic due to the fact
that 2 light masses and 1 heavy mass contribute along each crystallographic axis.

The anisotropy of the electronic effective mass {of} the conduction band can
be approximately taken into account in the following way. We use the averaged
inverse band mass
\begin{equation}
\frac{1}{\bar{m}_{b}}=\frac{1}{3}\left(  \frac{1}{m_{x}}+\frac{1}{m_{y}}%
+\frac{1}{m_{z}}\right)  \label{mb1}%
\end{equation}
and the density-of-states band mass
\begin{equation}
m_{D}=\left(  m_{x}m_{y}m_{z}\right)  ^{1/3}. \label{md}%
\end{equation}
{The mass }${m}_{D}$ {appears in the prefactor of the linear term of the
specific heat. Comparing the mass }$m_{D}${ obtained from the experimental
specific heat~\cite{Ambler66,phillips} with the mass }$\bar{m}_{b}${ obtained
using optical spectral weights \cite{VDM-PRL2008} reveals the mass ratio of
the heavy and light bands to be about 27.} The expression (\ref{mb1}) replaces
the bare mass $m_{b}$ in the optical conductivity (\ref{4}) and in the memory
function (\ref{MF}). The polarization function of the free electron gas
(\ref{P1}) is calculated with the density-of-states mass $m_{D}$ instead of
$m_{b}$. The band degeneracy is taken into account through the degeneracy
factor which is equal to 3, both in the polarization function and in the
normalization equation for the chemical potential.\ The reduction of the
polaron optical conductivity band due to screening with band degeneracy turns
out to be less significant than without band degeneracy.

\section{Theory and experiment \label{sec:results}}

\subsection{Material parameters}

Several experimental parameters characterizing SrTi$_{1-x}$Nb$_{x}$O$_{3}$ are
necessary for the calculation of the large-polaron optical conductivity (see,
e.g., Refs. \cite{JPCM2006,Gervais93}): the LO- and TO-phonon frequencies, the
electron band mass, and the electronic dielectric constant $\varepsilon
_{\infty}$.

The electronic dielectric constant can be obtained using reflectivity spectra
of SrTi$_{1-x}$Nb$_{x}$O$_{3}$. At $T=10$ K, the reflectivity of SrTi$_{1-x}%
$Nb$_{x}$O$_{3}$ is $R\approx0.16$ for $\Omega\approx5000$ cm$^{-1}$. The
electronic dielectric constant can be approximated using the expression%
\begin{equation}
R\left(  \Omega\right)  =\left\vert \frac{\sqrt{\varepsilon\left(
\Omega\right)  }-1}{\sqrt{\varepsilon\left(  \Omega\right)  }+1}\right\vert
^{2} \label{R}%
\end{equation}
and assuming that $\Omega=5000$ cm$^{-1}$ is a sufficiently high frequency to
characterize the electronic response. From (\ref{R}) it follows that for
SrTi$_{1-x}$Nb$_{x}$O$_{3}$, $\varepsilon_{\infty}\approx5.44$.

In order to determine the optical-phonon frequencies, we use {the}
experimental data from available sources \cite{VDM-PRL2008,Gervais93}. In Ref.
\cite{VDM-PRL2008}, three infrared active phonon modes are observed at room
temperature: at 11.0 meV, 21.8 meV and 67.6 meV. With decreasing temperature,
the lowest-frequency infrared-active phonon mode shows a strong red shift upon
cooling, and saturates at about 2.3 meV at 7 K. Those infrared-active phonon
modes are associated with the polar TO-phonons. The TO-phonon frequencies
determined in Ref. \cite{Gervais93} for SrTi$_{1-x}$Nb$_{x}$O$_{3}$ with
$x=0.9\%$ at $T=300$ K are 100 cm$^{-1}$, 175 cm$^{-1}$ and 550 cm$^{-1}$. The
corresponding TO-phonon energies are 12.4 meV, 21.7 meV and 68.2 meV.

Refs. \cite{VDM-PRL2008} and \cite{Gervais93} are used as sources for phonon
parameters. In Ref. \cite{Gervais93}, the TO-phonon frequencies are calculated
on the basis of reflectivity measurements using a model dielectric function to
fit experimental data. In Ref. \cite{VDM-PRL2008}, the TO-phonon frequencies
are obtained from an analysis of both reflectivity and transmission spectra,
using inversion of the Fresnel equations of reflection and transmission
coefficients and the Kramers-Kronig transformation of the reflectivity
spectra. The TO-phonon energies reported in Refs. \cite{VDM-PRL2008} and
\cite{Gervais93} are in close agreement. This confirms the reliability of both
experimental data sources \cite{VDM-PRL2008,Gervais93}. The values of the
TO-phonon frequencies used in our calculation are taken from the experiment
\cite{VDM-PRL2008} because they are directly related to the samples of
SrTi$_{1-x}$Nb$_{x}$O$_{3}$ for which the comparison of theory and experiment
is made in the present work.

\begin{table*}\centering
\caption{Optical-phonon frequencies and partial coupling constants of doped strontium titanate}%
\begin{tabular}
[c]{|l|l|l|l|l|l|l|l|l|}\hline
$x$ & $x=0.1\%$ & $x=0.1\%$ & $x=0.2\%$ & $x=0.2\%$ & $x=0.9\%$ & $x=0.9\%$ &
$x=2\%$ & $x=2\%$\\
$T$ & $T=7$ K & $T=300$ K & $T=7$ K & $T=300$ K & $T=7$ K & $T=300$ K & $T=7$
K & $T=300$ K\\\hline
$\hbar\omega_{T,1}$ (meV) & 2.27 & 11.5 & 2.63 & 11.5 & 6.01 & 12.1 & 8.51 &
13.0\\\hline
$\hbar\omega_{L,1}$ (meV) & 21.2 & 21.2 & 21.2 & 21.2 & 21.2 & 21.2 & 21.2 &
21.2\\\hline
$\alpha_{1}$ & 0.021 & 0.013 & 0.021 & 0.013 & 0.017 & 0.013 & 0.017 &
0.013\\\hline
$\hbar\omega_{T,2}$ (meV) & 21.2 & 21.8 & 21.2 & 21.8 & 21.2 & 21.8 & 21.2 &
21.8\\\hline
$\hbar\omega_{L,2}$ (meV) & 58.4 & 58.4 & 58.4 & 58.4 & 58.4 & 58.4 & 58.4 &
58.4\\\hline
$\alpha_{2}$ & 0.457 & 0.414 & 0.457 & 0.414 & 0.452 & 0.414 & 0.447 &
0.409\\\hline
$\hbar\omega_{T,3}$ (meV) & 67.6 & 67.1 & 67.6 & 67.1 & 67.6 & 67.1 & 67.6 &
67.1\\\hline
$\hbar\omega_{L,3}$ (meV) & 98.7 & 98.7 & 98.7 & 98.7 & 98.7 & 98.7 & 98.7 &
98.7\\\hline
$\alpha_{3}$ & 1.582 & 1.582 & 1.582 & 1.580 & 1.576 & 1.578 & 1.570 &
1.574\\\hline
$\alpha_{\mathrm{eff}}$ & 2.06 & 2.01 & 2.06 & 2.01 & 2.05 & 2.01 & 2.03 &
2.01\\\hline
\end{tabular}
\label{table1}
\end{table*}

The TO phonon frequencies from Ref. \cite{VDM-PRL2008} can be used when they
are complemented with corresponding LO phonon frequencies. However,
Ref.~\cite{VDM-PRL2008} does not contain data of the LO-phonon frequencies. In
the present calculation we use the LO phonon frequencies from
Ref.~\cite{Gervais93}.

The averaged band mass (\ref{mb1}) is taken to be $\bar{m}_{b}=0.81m_{e}$
(where $m_{e}$ is the electron mass in vacuum) according to experimental data
from Ref.~\cite{Comments2}. Using the ratio of the heavy mass ($m_{z}$) to the
light mass ($m_{x}=m_{y}$), $m_{z}/m_{x}=27$, we find the density-of states
band mass $m_{D}\approx1.65m_{e}.$

The TO- and LO- phonon frequencies and the resulting partial coupling
constants calculated using the mass $\bar{m}_{b}$ are presented in Table 1.%

The effective coupling constant in Table 1 is determined following Ref.
\cite{Ferro}, as a sum of partial coupling constants $\alpha_{j}$,%
\begin{equation}
\alpha_{\mathrm{eff}}\equiv\sum_{j}\alpha_{j} \label{aeff}%
\end{equation}
The result $\alpha_{\mathrm{eff}}\sim2$ shows that the electron-phonon
coupling strength in SrTi$_{1-x}$Nb$_{x}$O$_{3}$ lies in the {intermediate to
weak coupling range}, and the conditions for small polaron formation are not
fulfilled. This analysis indicates that the large-polaron picture -- rather
than the small-polaron description is suitable for the interpretation of the
mid-infrared band of the optical conductivity of SrTi$_{1-x}$Nb$_{x}$O$_{3}$.

We use the actual electron densities for the samples studied in
Ref.~\cite{VDM-PRL2008} based on the unit cell volume (59.5 cubic angstrom)
and the chemical composition ($x$ is the doping level). These carrier
densities (see Table 2) are confirmed by measurements of the Hall constants.%

\begin{table}[b] \centering
\caption{Electron densities of SrTi$_{1-x}$Nb$_{x}$O$_3$}
\begin{tabular}
[c]{|l|l|}\hline
$x\,(\%)$ & $n_{0}$ (cm$^{-3}$)\\\hline
0.1 & $1.7\times10^{19}$\\\hline
0.2 & $3.4\times10^{19}$\\\hline
0.9 & $1.5\times10^{20}$\\\hline
2.0 & $3.4\times10^{20}$\\\hline
\end{tabular}
\label{Table2}
\end{table}

\subsection{Optical conductivity spectra}

We calculate the large-polaron optical conductivity spectra for SrTi$_{1-x}%
$Nb$_{x}$O$_{3}$ using the approach of Ref. \cite{TD2001} as adapted in Ref.
\cite{draft} to take into account multiple LO-phonon branches. We also include
in the numerical calculation the TO-phonon contribution to the optical
conductivity, described by an oscillatory-like model dielectric function (see,
e.g., Ref. \cite{Gervais93}):%
\begin{equation}
\operatorname{Re}\sigma_{TO}\left(  \Omega\right)  =\sum_{j}\sigma_{0,j}%
\frac{\gamma_{j}^{2}}{\left(  \Omega-\omega_{T,j}\right)  ^{2}+\gamma_{j}^{2}%
}, \label{sto}%
\end{equation}
where the weight coefficients $\sigma_{0,j}$ and the damping parameters
$\gamma_{j}$ for each $j$-th TO-phonon branch are extracted from the
experimental optical conductivity spectra of Ref. \cite{VDM-PRL2008}. The
polaron-and the TO-phonon optical responses are treated as independent {of}
each other. Consequently the polaron-(\ref{4}) and TO-phonon (\ref{sto})
contributions enter the optical conductivity additively.

\begin{figure*}
\hspace{-1.7mm}\includegraphics[width=14cm]{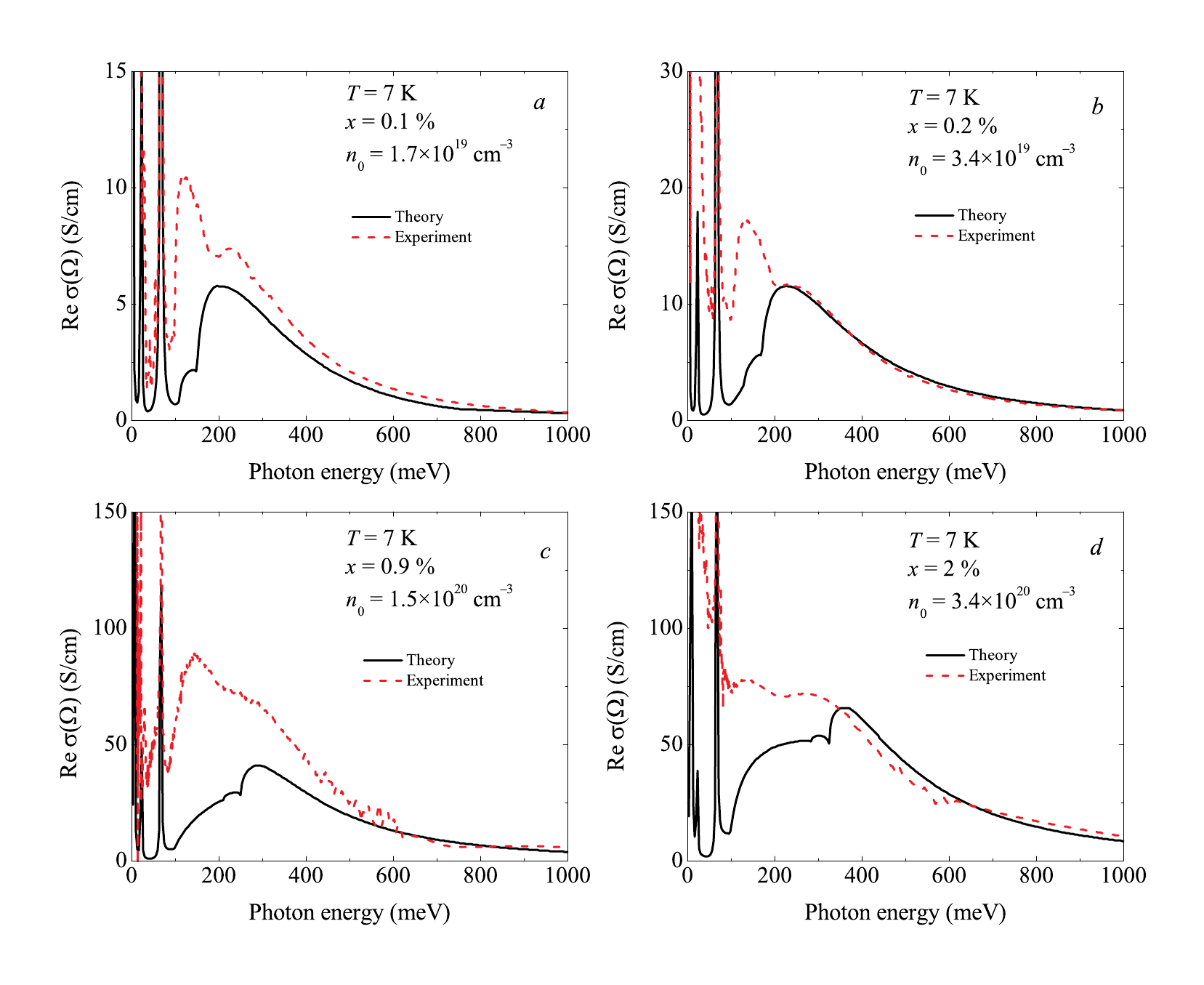}
\caption{The many-large-polaron optical conductivity compared
with the experiment~\cite{VDM-PRL2008} at $T=7$ K. The doping level is
$x=0.1\%$ (\emph{a}), 0.2{\%} (\emph{b}), 0.9\% (\emph{c}) and 2\%(\emph{d}).}
\label{fig_swtransfer}
\end{figure*}

{Following the procedure described above using the material parameters
discussed above, we obtain the theoretical large-polaron optical conductivity
spectra of SrTi$_{1-x}$Nb$_{x}$O$_{3}$ shown in Fig.~2 and Fig.~3 at 7 K and
300 K, respectively. In each graph also the experimental optical conductivity
spectra of Ref.~\cite{VDM-PRL2008} are shown. It should be emphasized that}
\emph{in the present calculation, there is no fitting of material constants
for the polaron contribution to }$\operatorname{Re}\sigma\left(
\Omega\right)  $\emph{. Even the magnitude of the optical conductivity, which
is often arbitrarily scaled in the literature, follows from first principles}.

At 7 K, the calculated optical conductivity based on the Fr\"{o}hlich model
and extended for a gas of large polarons as described in the present paper,
shows convincing agreement with the behavior of the experimental optical
conductivity for the high energy part of the spectra, i.e., $\hbar
\Omega\gtrapprox300$ meV. The experimental polaron optical conductivity of
SrTi$_{1-x}$Nb$_{x}$O$_{3}$ falls down at high frequencies following the power
law (derived in the present work and typical for large polarons) rather than
as a Gaussian exponent that would follow from the small-polaron theory. At
lower photon energies $\hbar\Omega\lessapprox200$ meV, the experiment shows
distinct peaks that are not explained within the polaron theory. They can be
due to other scattering mechanisms as discussed below.

\begin{figure*}
\hspace{-1.7mm}\includegraphics[width=14cm]{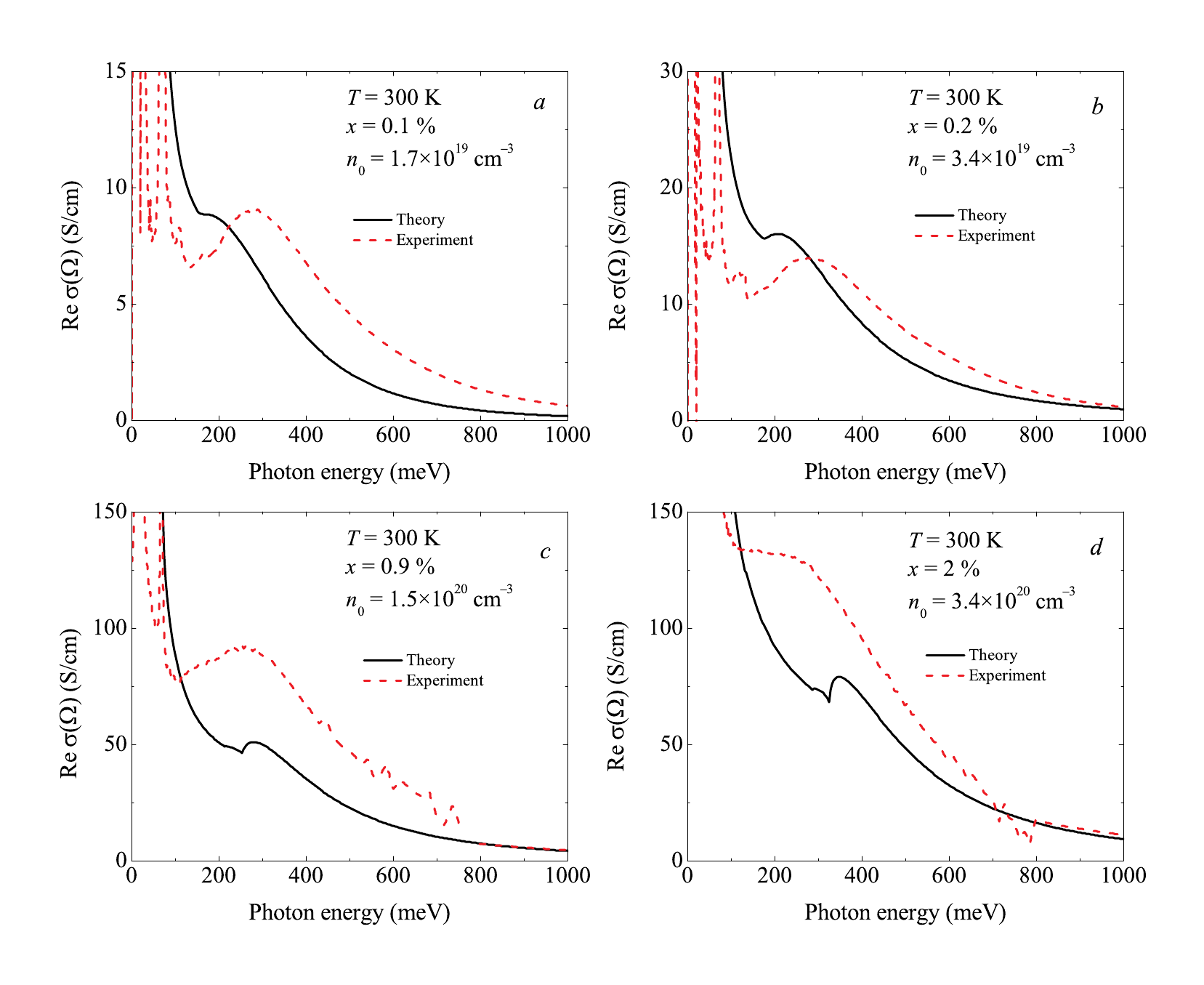}
\caption{The many-large-polaron optical conductivity compared
with the experiment~\cite{VDM-PRL2008} at $T=300$ K. The doping level is
$x=0.1\%$ (\emph{a}), 0.2{\%} (\emph{b}), 0.9\% (\emph{c}) and 2\% (\emph{d}).}
\label{fig_swtransfer}
\end{figure*}

The minor deviations between theoretical and experimental $\operatorname{Re}%
\sigma\left(  \Omega\right)  $ in the frequency range $\hbar\Omega
\gtrapprox300$ meV may be attributed to the difference between the actual
electron densities and the densities calculated on the basis of the unit cell
volume and the chemical composition. However, we prefer not to fit of the density.

The optical conductivity calculated for a single large-polaron
absorption~\cite{DSG} predicts an intensity 3-4 times larger than the
experimental data for the lowest doping level $x=0.1\%$, and therefore cannot
explain those data. For higher dopings, the overestimation of the magnitude of
the optical conductivity within the single-polaron theory is even larger than
for $x=0.1\%$. Therefore the many-polaron approach, used in the present work,
is essential.

At 300 K, in Fig.~3~(\emph{a},\ \emph{b\ },\emph{d}), the agreement between
theory and experiment is qualitative. Both experimental and theoretical
spectra show a maximum at the room-temperature optical conductivity spectra in
the range $\hbar\Omega\sim250$ meV. For the doping level $x=0.9\%$ the
calculated optical conductivity spectrum underestimates the experimental data,
as also observed at 7 K.

Many-body effects considerably influence the optical conductivity spectra of a
polaron gas. First, features related to the emission of a plasmon together
with a LO phonon \cite{TD2001} are manifested in the optical conductivity
spectra of the many-polaron gas at $T=7$ K as separate peaks whose positions
shift to higher energies with increasing doping level. At room temperature,
those peaks are strongly broadened and smoothened, and only a broad plasmon
feature is apparent. Second, the mid-infrared optical conductivity (per
particle) in SrTi$_{1-x}$Nb$_{x}$O$_{3}$ is decreasing at higher doping levels
due to the screening of the polar interactions, which is accounted for in the
present approach in which $S\left(  \mathbf{q},t\right)  $ is based on RPA.
The effect of screening can be illustrated by the fact that for $n_{0}%
\sim10^{20}$ cm$^{-3}$, the many-polaron optical conductivity \emph{per
particle} is reduced by about an order of magnitude compared to the
single-polaron optical conductivity. The reduction in intensity of the polaron
optical conductivity band can be interpreted as a decrease of the overall
electron-phonon coupling strength due to many-body effects. Correspondingly,
at high doping levels, the polaron mass $m^{\ast}$, determined by the sum rule
introduced in Ref. \cite{DLR77}%
\begin{equation}
\frac{\pi e^{2}n_{0}}{2m^{\ast}}+\int_{\omega_{L}}^{\infty}\operatorname{Re}%
\left(  \Omega\right)  d\Omega=\frac{\pi e^{2}n_{0}}{2\bar{m}_{b}}\label{sr}%
\end{equation}
is reduced, compared to the single-polaron effective mass. As shown in Refs.
\cite{TD2001,TD2001a}, the sum rule \cite{DLR77} remains valid for an
interacting polaron gas.

The large-polaron theory of the optical absorption based on Ref. \cite{TD2001}
explains without any fitting parameters the main characteristics and trends of
the observed spectra of Ref. \cite{VDM-PRL2008} in SrTi$_{1-x}$Nb$_{x}$O$_{3}%
$, including doping- and temperature dependence. Nevertheless, some features
of the experimental spectra remain to be explained. In particular, at $T=7$ K,
the pronounced peak at $\hbar\Omega\sim130$ meV in the experimental optical
conductivity is not accounted for by the present theoretical analysis. In the
theoretical spectra, peaks of much smaller intensity appear at about the same
frequency. In the large-polaron theory, those peaks are provided by the
interaction between electrons and the LO-phonon branch with energy
$\hbar\omega_{L,2}\approx58.4$ meV, accompanied by the emission of a plasmon
as described in Ref. \cite{TD2001}.

The intensity of the experimentally observed absorption peak at $\hbar
\Omega\sim130$ meV is considerably higher than described by the large-polaron
theory. In the low density limit, the experimental optical data more rapidly
approach the single polaron limit~\cite{DSG} than the theoretical predictions
based on Eq.~(\ref{xi}). This absorption peak at $\hbar\Omega\sim130$ meV may
be provided by other mechanisms, not controlled in the present study. E. g.,
electron-phonon interaction with low-frequency non-polar (e. g., acoustic)
phonons may contribute to the optical conductivity. The squared modulus
$\left\vert V_{q}\right\vert ^{2},$ which characterizes the coupling strength,
for the deformation electron-phonon interaction is $\left\vert V_{\mathbf{q}%
}\right\vert ^{2}\propto q$ \cite{PD1985}, while for the Fr\"{o}hlich
interaction, $\left\vert V_{\mathbf{q}}\right\vert ^{2}\propto q^{-2}$.
Consequently, for the deformation electron-phonon interaction, the
short-wavelength phonons may provide non-negligible contributions to the
optical conductivity. Also, at sufficiently large $q,$ Umklapp scattering
processes with acoustic phonons can play a role. The treatment of
contributions due to acoustic phonons (and other mechanisms) is the subject of
the future work. Another possible explanation of the absorption peak at
$\hbar\Omega\sim130$ meV is weakened screening in the corresponding energy
range due to dynamical-exchange \cite{BDL}.

\section{Conclusions \label{sec:conclusions}}

Many-polaron optical conductivity spectra, calculated (based on Ref.
\cite{TD2001}) within the large-polaron picture without adjustment of material
constants, explain essential characteristics of the experimental optical
conductivity~\cite{VDM-PRL2008}. The intensities of the calculated
many-polaron optical conductivity spectra and the intensities of the
experimental mid-infrared bands of the optical conductivity spectra of
SrTi$_{1-x}$Nb$_{x}$O$_{3}$ (from Ref.~\cite{VDM-PRL2008}) are comparable for
all considered values of the doping parameter. The doping dependence of the
intensity of the mid-infrared band in the theoretical large-polaron spectra is
similar to that of the experimental data of Ref.~\cite{VDM-PRL2008}. In the
high-frequency range, the theoretical absorption curves describe well the
experimental data (especially at low temperature). A remarkable difference
between the present theoretical approach and experiment is manifested on the
low frequency side of the mid-infrared range, where the experimental optical
conductivity shows a sharp and pronounced peak for $\hbar\Omega\sim130$ meV at
7 K. Although the theoretical curve also shows a feature around the same
frequency, its intensity is clearly underestimated. This peak in the
absorption spectrum at $\hbar\Omega\sim130$ meV remains to be explained. The
value of the effective electron-phonon coupling constant obtained in the
present work ($\alpha_{eff}\approx2$) corresponds to the intermediate coupling
strength of the large-polaron theory.

The alternative small-polaron and mixed-polaron models for the optical
conductivity require several fitting parameters. Furthermore, we find that the
mixed-polaron model would need a major adjustment of the overall intensity in
order to fit experimental spectra. 

Contrary to the case of the large polaron, the small-polaron parameters cannot
be extracted from experimental data. Moreover, the small-polaron model, for
any realistic choice of parameters, shows a frequency dependence in the
high-frequency range which is different from that of the experimental optical
conductivity. Both the experimental and the theoretical large-polaron optical
conductivity decrease as a power function at high frequencies, while the
small-polaron optical conductivity falls down exponentially for sufficiently
high $\Omega$.

In summary, the many-body large-polaron model based on the Fr\"{o}hlich
interaction accounts for the essential characteristics (except --
interestingly -- for the intensity of a prominent peak at $\hbar\Omega\sim130$
meV, that constitutes an interesting challenge for theory) of the experimental
mid-infrared optical conductivity band in SrTi$_{1-x}$Nb$_{x}$O$_{3}$ without
any adjustment of material parameters. The large-polaron model gives then a
convincing interpretation of the experimentally observed mid-infrared band of
SrTi$_{1-x}$Nb$_{x}$O$_{3}$.
\vspace{7mm}
\begin{acknowledgments}
\label{sec:acknowledge} This work was supported by FWO-V projects G.0356.06,
G.0370.09N, G.0180.09N, G.0365.08, the WOG WO.035.04N (Belgium), and by the
Swiss National Science Foundation under Grant No. 200020-125248 and the
National Center of Competence in Research (NCCR) Materials with Novel
Electronic Properties--MaNEP.
\end{acknowledgments}









\end{document}